\documentstyle[prabib,aps,preprint,epsf]{revtex}

\ifx\undefined\ifEPRINT
\newcommand{\GETPS}[2]{} 				% Without Figures

\else
\newcommand{\GETPS}[2]{\epsfxsize=#1in\epsfbox{#2}}	% With Figures

\fi

\date{}
\begin{document}

\noindent
%%%%%%%%%%%%%%%%%%%%%%%%%%%%%%%%
\begin{center}{\Large \bf 
Triaxial deformation in $^{10}$Be} \\

\vskip.25in
{\it N. Itagaki,$^1$\footnote{E-mail: itagaki@phys.s.u-tokyo.ac.jp}
S. Hirose,$^1$ T. Otsuka,$^{1,2}$ S. Okabe,$^3$ and K. Ikeda$^2$}

{$^1$ \it 
Department of physics, University of Tokyo, Hongo, Tokyo 113-0033, Japan
}

{$^2$ \it 
The Institute of Physical and Chemical Research (RIKEN),
Wako, Saitama 351-0198, Japan}

{$^3$ \it Center for Information and Multimedia Studies,
Hokkaido University, Sapporo 060-0810, Japan}

\end{center}

\begin{abstract}                % DON'T CHANGE THIS LINE
The triaxial deformation in $^{10}$Be is investigated 
using a microscopic $\alpha$+$\alpha$+$n$+$n$ model.
The states of two valence neutrons are classified 
based on the molecular-orbit (MO) model, and the $\pi$-orbit
is introduced about the axis connecting
the two $\alpha$-clusters for the description
of the rotational bands.
There appear two rotational bands
comprised mainly 
of $K^\pi = 0^+$ and $K^\pi = 2^+$, respectively,
at low excitation energy, 
where the two valence neutrons occupy $K^\pi = 3/2^-$ or 
$K^\pi = 1/2^-$ orbits.
The triaxiality and the $K$-mixing are discussed
in connection to the molecular structure, particularly,
to the spin-orbit splitting.
The extent of the triaxial deformation is evaluated in terms of
the electro-magnetic transition matrix elements
(Davydov-Filippov model, $Q$-invariant model),
and density distribution 
in the intrinsic frame.
The obtained values turned out to be
$\gamma = 15^o \sim 20^o$.
\end{abstract}
\begin{center}
PACS number(s): 21.10.-k, 21.60.Gx
\end{center}

\section{INTRODUCTION}
Recently, exotic structures of the Be isotopes 
have been theoretically and experimentally studied,
and many new phenomena have been discussed
\cite{Iwasaki1,Iwasaki2,Navin,Kor,RB,Freer1,Freer2}.
One of them is the appearance of the 
cluster rotational band structure in the excited states
of $^{10}$Be ($\alpha$+$^6$He)\cite{RB,Freer2} 
and $^{12}$Be ($\alpha$+$^8$He, $^6$He+$^6$He)\cite{Kor,Freer1}. 
We have suggested \cite{Ita,Ita2,Ita3}, 
based on the molecular-orbit (MO) model,
that the development of $\alpha$-$\alpha$ cluster structure depends on 
the neutron orbits located around the core comprised of $\alpha$-clusters.
In the second $0^+$ state of $^{10}$Be, an anomalously prolonged
$\alpha$-$\alpha$ clustering structure emerges due to the valence neutrons
located along the $\alpha$-$\alpha$ axis\cite{Ita,Ita2}.

The nucleus
$^{10}$Be has been known to have a strong $\beta$ deformation
due to the $\alpha$-$\alpha$ core.
In this paper, we discuss another aspect, a triaxial deformation.
The triaxial deformation is possible as a result of 
the dynamics of the two valence neutrons.
The triaxiality of $^{10}$Be 
has been theoretically discussed
based on the deformed oscillator model
\cite{Harvey}, in which a $\gamma$ distortion of 34.8$^o$ has been predicted.
According to Davydov-Filippov model\cite{Davydov},
this $\gamma$ value suggests that 
the excitation energy of 
the second $2^+$ state is 2.3 times
higher than that of the first $2^+$ state.
In $^{10}$Be, the first $2^+$ state is observed at 3.358 MeV,
and several $2^+$ states have been observed around 8 MeV region.
Because of the presence of the $2^+$ states in this energy region,
recently, an experimental study has been performed in order to
identify a triaxial 
structure of $^{10}$Be\cite{Curtis}.

In this paper, we discuss the triaxiality of $^{10}$Be 
as a composite system of the $\alpha$-$\alpha$ core and two valence neutrons,
comparing with other theoretical models. 
In our previous study on $^{10}$Be\cite{Ita}, 
all of the observed low-lying positive- and 
negative-parity states are explained as combinations of three
basic orbits ($K^\pi = 3/2^-$, $1/2^+$, and $1/2^-$)
of two valence neutrons around the two $\alpha$-clusters.
Here, the $z$-axis is taken to be the axis connecting two $\alpha$-clusters.
If we adopt $K^\pi = 3/2^-$ or $1/2^-$ orbits for the 
two valence neutrons,
there appear two rotational bands in low energy;
one is dominated by the $K = 0$ intrinsic structure
and the other by $K = 2$.
The calculated two $2^+$ states of these bands can be related
to the observed first $2^+$ state at 3.358 MeV and 
the second $2^+$ state at 5.958 MeV.
It is therefore important to show how 
the triaxial intrinsic configuration emerges in 
these states, and how
the orbits of the valence neutrons
deviate from the axial symmetry.
Here, we calculate the electro-magnetic transition
between these $K = 0$ and $K = 2$ bands 
(B(E2: $K = 2 \to K = 0 $))
as a signal of a triaxial deformation.
This transition is suppressed when the orbitals are of pure
axial symmetry.
However, the orbitals may deviate from the axial symmetry, 
when the valence neutrons are mutually more correlated and
form a localized di-neutron pair
due to the neutron-neutron interaction.
The recoil effect of the valence neutrons with
respect to the $\alpha$-$\alpha$ core 
then plays a role to break the axial symmetry of the charge distribution.

This paper is organized as follows. In section II, we give
a description of the single-particle orbits around 
the two $\alpha$-clusters based on MO.
In section III, we show our results on $^{10}$Be,
and a conclusion is given in section IV.
%%%%%%%%%%%%%%%%%%%%%%%%%%%%%%%%%%%%%%%%%%%%%

%%%%%%%%%%%%%%%%%%%%%%%%%%%%%%%%%%%%%%%%%%%%%%%%%%%%%
%%%%%%	Model and Method	%%%%%
\section{Extended molecular-orbit model}
\subsection{Wave function of $^{10}$Be}
We introduce a microscopic $\alpha$+$\alpha$+$2n$ model for $^{10}$Be. 
The neutron configurations are introduced
based on the molecular orbit (MO) picture\cite{Abe,Okabe-S,Seya}.
The total wave function is fully antisymmetrized and
expressed as a superposition of Slater determinants
with various configurations of the valence neutrons. 
The Slater determinants are also superposed with respect to
different relative distances between the two $\alpha$ clusters ($d$).
The projection
to the eigen states of angular momentum $J$ is numerically performed.
All nucleons are described by Gaussians 
with a common
oscillator parameter $s = {1 \over \sqrt{2 \nu}}$ set equal to 1.46 fm.
The $\alpha$ cluster located at $R_\alpha$ 
on the $z$-axis contains four nucleons.
\begin{equation}
\phi^{(\alpha)}=
G_{R_\alpha}^{p\uparrow}
G_{R_\alpha}^{p\downarrow}
G_{R_\alpha}^{n\uparrow}
G_{R_\alpha}^{n\downarrow}
\chi_{p\uparrow}
\chi_{p\downarrow}
\chi_{n\uparrow}
\chi_{n\downarrow}.
\end{equation}
Here, $G$ represents a Gaussian:
\begin{equation}
G_{R_\alpha} = \left( {2\nu \over \pi}\right)^{3 \over 4}
\exp[-\nu(\vec r-\vec R_\alpha)^2],
\ \ \ \ \ \ \nu=1/2s^2, 
\end{equation}
where $\vec R_\alpha = R_\alpha \vec e_z$
with $\vec e_z$ being the unit vector pointing the $z$-direction.
The actual value of $R_\alpha$ is $d/2$ or $-d/2$ in this work.
The wave function of the $i-$th valence neutron ($\phi_{i}\chi_{i}$)
is expressed by a linear combination of local Gaussians:
\begin{equation}
\phi_i \chi_i = \sum_j g_j G_{R_j} \chi_i,
\end{equation}
\begin{equation}
G_{R_j}=\left( {2\nu \over \pi}\right)^{3 \over 4}
\exp[-\nu(\vec r_i-\vec R_j)^2],
\end{equation}
where $\vec R_j$ is
a parameter corresponding to the Gaussian center.

The two valence neutrons are classified according to the MO picture.
In the MO model, the states of the valence neutrons are expressed 
by a linear combination of orbits 
around two $\alpha$ clusters.
Here, we notice that the valence neutrons and the neutrons in
the $\alpha$ clusters are identical particles,
and the valence neutrons must be orthogonal to the
forbidden space due to the $\alpha$-$\alpha$ core.
The lowest orbit then
has one node and negative parity, that is the $p$-orbit.
We take the harmonic-oscillator-type wave function for the $p$-orbit.
Such $p$-orbits in the $x$-, $y$-, and $z$-direction are denoted as
$\psi_x$, $\psi_y$, and $\psi_z$, respectively.
We now classify single particle orbits 
for the valence neutrons 
in terms of $K$ quantum numbers:
\begin{equation}
\psi_x + i\psi_y = (x+iy)\exp[-\nu r^2] \propto rY_{11} \exp[-\nu r^2],
\end{equation}
\begin{equation}
\psi_z = z\exp[-\nu r^2] \propto rY_{10} \exp[-\nu r^2],
\end{equation}
\begin{equation}
\psi_x - i\psi_y = (x-iy)\exp[-\nu r^2] \propto rY_{1-1} \exp[-\nu r^2].
\end{equation}
These $p$-orbits have $K = 1$, 0, and $-1$, respectively.
Here, $x$, $y$, and $z$ are relative to $\vec R_\alpha$.

Now we construct MO from these $p$-orbits. 
Since each valence neutron
can move around one of the two $\alpha$-clusters,
there should be two sets of the $p$-orbits defined with
respect to those two $\alpha$-clusters.
Thus, the $(\psi_x \pm i\psi_y)$ orbit whose center is 
shifted at $+a$ ($-a$) on the $z$-axis
is denoted as $(\psi_x \pm i\psi_y)_{+a}$ ($(\psi_x \pm i\psi_y)_{-a}$).
As linear combinations of these orbits,
the lowest MOs are expressed as:
\begin{equation}
\psi^{MO}_1 = (\psi_x+i\psi_y)_{+a}+(\psi_x+i\psi_y)_{-a},
\end{equation}
\begin{equation}
\psi^{MO}_{-1} = (\psi_x-i\psi_y)_{+a}+(\psi_x-i\psi_y)_{-a}.
\end{equation}
These orbits clearly have $K = 1 $ and $K = -1$, respectively.
In these cases, the classical picture of $p$-orbit 
is a circular motion about the $\alpha$-$\alpha$ $(z)$ axis. 
This is the so-called $\pi$-orbit. 

If we take a linear combination of $(\psi_z)_{+a}$ and $(\psi_z)_{-a}$,
the orbit becomes the so-called $\sigma$-orbit,
just along the $\alpha$-$\alpha$ ($z$) axis. 
This orbit is a higher-nordal orbit and only relevant
to the second $0^+$ state, discussed also in 
antisymmetrized molecular dynamics (AMD)
calculations\cite{EnyoBe10}
and in stochastic variational method (SVM)
calculations\cite{Ogawa},
and now we disregard this orbit.

These 
$(\psi_x)_{+a}$, $(\psi_x)_{-a}$, $(\psi_y)_{+a}$, and $(\psi_y)_{-a}$
orbits can be approximated by a
combination of two local Gaussians, whose centers are  
shifted by a variational parameter $b$ perpendicular to the $z$-axis.
We thus use the following wave functions,
\begin{equation}
(\psi_x)_{+a} \propto G_{a\vec e_z+b\vec e_x}-G_{a\vec e_z-b\vec e_x}, \ \ \
(\psi_y)_{+a} \propto G_{a\vec e_z+b\vec e_y}-G_{a\vec e_z-b\vec e_y}, \ \ \
\cdot \cdot \cdot \cdot.
\end{equation}
The values of these 
parameters $a$ and $b$  are variationally determined by using 
the cooling method in AMD \cite{Enyo95,Enyo95b,Enyo95c,AMD,Ono92}
independently for each $\alpha$-$\alpha$ distance.
Since the rotational symmetry about the $z$-axis 
is broken with Eq. (10), 
the wave functions do not have exactly conserved quantum number $K$.
On the other hand, the parameter $b$ is small enough usually,
and the $K^\pi$ number is preserved to a good extent, and is
expressed hereafter as $\bar K^\pi$.

When these orbital part of 
the wave functions are coupled with the spin part,
the orbits with $\bar K^\pi = 3/2$, $1/2$, $-1/2$, and $-3/2$
are introduced, as follows:
\begin{equation}
|3/2^- \rangle = \psi^{MO}_1   |n\uparrow \rangle, \ \ \
|1/2^- \rangle = \psi^{MO}_1   |n\downarrow \rangle, 
\end{equation}
and their time reversal,
\begin{equation}
|-3/2^- \rangle = \psi^{MO}_{-1}   |n\downarrow \rangle, \ \ \
|-1/2^- \rangle = \psi^{MO}_{-1}   |n\uparrow \rangle. 
\end{equation}

\subsection{Hamiltonian}
The Hamiltonian consists of the kinetic energy term,
the central two-body interaction term, the two-body spin-orbit 
interaction term,
and the Coulomb interaction term: 
\begin{equation}
{\cal H}=\sum_i T_i - T_{cm} + \sum_{i < j} V_{ij}   
+ \sum_{i < j} V^{ls}_{ij} 
+ \sum_{i < j} {e^2 \over 4r_{ij}} (1-\tau_z^i)(1-\tau_z^j).
\end{equation}
The effective nucleon-nucleon interactions
are Volkov No.2\cite{VolkovInt} for the central part
and the G3RS spin-orbit term\cite{G3RS} 
for the spin-orbit part, as follows:
\begin{equation}
V_{ij} = \{V_1 e^{-a_1r_{ij}^2} + V_2 e^{-a_2r_{ij}^2}\} 
\{ W-MP^{\sigma}P^{\tau}+BP^{\sigma}-HP^{\tau} \},
\end{equation}
\begin{equation}
V^{ls}_{ij} = V^{ls}_0 \{ e^{-a_1r_{ij}^2}-e^{-a_2r_{ij}^2}\} 
{\vec L \cdot \vec S}P_{31},
\end{equation}
where $P_{31}$ is a projection
operator onto the triplet odd state, 
and $\vec L$ and $\vec S$ operators represent the relative angular momentum
and total spin of the interacting two nucleons, respectively.
The parameters are
$V_1 = -60.650$ MeV, $V_2 = 61.140$ MeV, $a_1 = 0.309$ fm$^{-2}$ 
and $a_2 = 0.980 $ fm$^{-2}$ 
for the central interaction,
and $V^{ls}_0$ = 2000 MeV, 
$a_1 = 5.00$ fm$^{-2}$, and $a_2 = 2.778$ fm$^{-2}$
for the spin-orbit interaction.
We employ
the Majorana exchange parameter $M = 0.6$ ($W = 0.4$),
the Bartlett exchange parameter $B = 0.125$
and the Heisenberg exchange parameter $H = 0.125$ for the Volkov
interaction
(using $B$ and $H$, there is no neutron-neutron bound state).
All these parameters are determined 
from $\alpha+n$ and $\alpha+\alpha$ 
scattering phase shifts and the binding energy of
the deuteron\cite{Okabe79}.

\subsection{Configurations of the valence neutrons}
For $^{10}$Be, we introduce three configurations of 
$\Phi((3/2^-)^2)$, $\Phi((1/2^-)^2)$, and 
$\Phi(3/2^-\ 1/2^-)$ for the two valence neutrons,
and mixing amplitude of these configurations are variationally
determined after the angular momentum projection.

It will be shown that
the $\Phi((3/2^-)^2)$ configuration,
where valence neutrons occupy orbits with 
$\bar K^\pi = 3/2^-$ and $\bar K^\pi = -3/2^-$
is the main component of the ground state.
The energy shift due to the spin-orbit interaction
acting effectively on a valence neutron 
is proportional to $- \langle \vec l \cdot \vec s \rangle$.
It is, therefore, favored that
the orbital part and the spin part of $\bar K$ 
are coupled to be parallel.
This means that the spin-up valence neutron ($|n\uparrow \rangle$) 
has $\bar K^{\pi}=3/2^-$,
while the spin-down valence neutron ($|n\downarrow \rangle$) 
$\bar K^{\pi}=-3/2^-$, giving rise to the wave function of the 
two valence neutrons:
\begin{equation}
\Phi((3/2^-)^2) 
={\cal A}[\phi^{(\alpha)}_1\phi^{(\alpha)}_2
(\phi_{c1}\chi_{c1})(\phi_{c2}\chi_{c2})],
\end{equation}
with
\begin{equation}
\phi_{c1}\chi_{c1}= \psi^{MO}_1   |n\uparrow \rangle, \ \ \
\phi_{c2}\chi_{c2}= \psi^{MO}_{-1}|n\downarrow \rangle.
\end{equation}

We also introduce a basis state $\Phi((1/2^-)^2)$,
which the spin-orbit interaction does not favor.
They are defined by flipping the spin-part of 
the $3/2^-$ and $-3/2^-$ wave functions.
and the valence neutrons occupy orbits of
$\bar K^\pi = 1/2^-$ and $\bar K^\pi = -1/2^-$:
\begin{equation}
\phi_{c1}\chi_{c1}= \psi^{MO}_1   |n\downarrow \rangle, \ \ \
\phi_{c2}\chi_{c2}= \psi^{MO}_{-1}|n\uparrow \rangle.
\end{equation}

The $\Phi(3/2^-\ 1/2^-)$ configuration with $\bar K = 2$
is constructed as a combination of the valence neutrons in the 
$\bar K^\pi = 3/2^-$ and $\bar K^\pi = 1/2^-$ orbits.
\begin{equation}
\phi_{c1}\chi_{c1}= \psi^{MO}_1   |n\uparrow \rangle, \ \ \
\phi_{c2}\chi_{c2}= \psi^{MO}_1   |n\downarrow \rangle.
\end{equation}
%
%%%%%%%%%%%%%%%%%%%%%%%%%%%%%%%%%%%%%%%%%%%%%%%%%%%%%
\section{Triaxial structure in $^{10}$Be}
In $^{10}$Be, we adopt three configurations
$\Phi((3/2^-)^2)$, $\Phi((1/2^-)^2)$, and 
$\Phi(3/2^-\ 1/2^-)$,
where single particle orbits with $\bar K = 3/2^-, 1/2^-$
have been introduced.
The values of the parameters 
for the valence neutrons are obtained variationally.
The parameter $a$ describes the positions of the Gaussian centers
on the $\alpha$-$\alpha$ ($z$) axis, and parameter 
$b$ corresponds to the rotation radius of the $\pi$-orbit
about the $\alpha$-$\alpha$ axis.
The optimized values are listed in Table I.

\noindent
\begin{center}
--------------\\
Table I \\
--------------\\
\end{center}
The parameter $a$ is obtained
to be approximately the same value as $d/2$.
Therefore, the picture ``orbit around the $\alpha$-cluster''
works well, while there is a small deviation 
of the orbit along the $z$-axis.
The parameter $b$
depends on the configurations.
For $\Phi((3/2^-)^2)$,
$b$ appears to be $\sim$1 fm.
For $\Phi((1/2^-)^2)$, $b$
becomes longer as being 2 fm,
since the spin-orbit interaction acts
attractively for $\Phi((3/2^-)^2)$ 
whereas repulsively for $\Phi((1/2^-)^2)$. 

Before performing the angular momentum projection, we discuss
the triaxial deformation of the intrinsic wave function.
The intrinsic density of $\Phi((3/2^-)^2)$ 
is shown in Fig. 1 ((a): $xz$-plane, (b): $xy$-plane), 
where the $\alpha$-$\alpha$ distance is chosen to be optimal,
as it turns out to be 3 fm.

\noindent
\begin{center}
------------------\\
Fig. 1 (a), (b)\\
------------------\\
\end{center}
The intrinsic density is defined as a snap-shot of a rotating system.
The plane in which the spin-up valence neutrons 
stays at a given moment is defined as the $xz$-plane, with the
$z$-axis being the $\alpha$-$\alpha$ axis.
Therefore, the spin-up valence neutron is fixed on the $xz$-plane,
and the spin-down valence neutron occupy the normal $\bar K = -3/2^-$ orbit.
The intrinsic density on the $xz$-plane (Fig.1 (a))
shows a large $\beta$ deformation along the $z$-axis
due to the presence of two $\alpha$-clusters.
As for the $xy$-plane (Fig. 1 (b)),
the density shows deviation from circular distribution
and mixture of a triaxial component is evident.
The contour-line of 0.01 in Fig. 1 (a) and (b)
suggests the deformation of 
this nucleus is $\beta \sim0.5 $ and $\gamma = 11^o$.

We then perform the angular momentum projection,
where the wave function is rotated and integrated over the Euler angle,
and the rotational symmetry is restored.
Here, the $K$-mixing between $K = 0 $ and $K = 2 $
finally determines the degree of the triaxiality of the system.  

This intrinsic wave function is
numerically projected to the eigen states of angular momentum,
and the basis states with parameters listed in Table I are 
superposed by generator coordinate method (GCM).
Here, the coefficients representing 
linear combinations of Gaussians for each single particle orbit
are treated as variational parameters
to take into account deviations from the original orbits.
Because of this, not only ``single-particle'' MO state 
where the valence neutrons independently rotate around 
the $\alpha$-$\alpha$ core, but also more complex states
where the valence neutrons are 
mutually more correlated as a di-neutron cluster can be included. 

The energy levels of $^{10}$Be are shown in Fig. 2. 
Using our framework, the binding energy of one $\alpha$-cluster
is calculated to be 27.5 MeV, and the threshold energy 
of free $\alpha$+$\alpha$+$n$+$n$ system
is $2 \times (-27.5) = -55.0$ MeV.
Experimentally, the ground state is lower than
this energy by 8.4 MeV.

\noindent
\begin{center}
--------------\\
Fig. 2 \\
--------------\\
\end{center}
The left and middle columns in Fig. 2 show the calculated energy levels
with the constraints of $K = 0$ and $K = 2$, respectively,
and the right column shows the result after the $K$-mixing.
In $K = 0$ column, it can be seen that the ground $0^+$ state at $-60.5$ MeV,
the $2^+$ state at $-56.9$ MeV, and the $4^+$ state at $-47.6$ MeV
form a rotational band structure. 
They fit quite well into the $J(J+1)$ rule,
and the dominant component is $(3/2^-)^2$ for the two valence neutrons.
For $K = 2$ column, 
the $2^+$ state at $-54.9$ MeV, 
the $3^+$ state at $-51.1$ MeV, 
and the $4^+$ state at $-46.1$ MeV 
form a rotational band structure. 
They also fit quite well into the $J(J+1)$ rule.
The $K = 2 $ band dominantly has a component of $(3/2^-)(1/2^-)$ 
for the two valence neutrons,
where one of them feels the spin-orbit interaction attractively and
the other feels repulsively.
After the mixing of these two bands, the $2^+$ state with $K = 0$
at $-56.9$ MeV becomes $-57.3$ MeV
($E_x = 3.1$ MeV, experimentally 3.358 MeV), 
and $2^+$ state with $K = 2$
is slightly pushed up to $-54.9$ MeV
($E_x = 5.6$ MeV, experimentally 5.958 MeV). 
The first $2^+$ state has the squared overlap with the $K = 0$
state by 0.96, and the second $2^+$ state has the squared overlap
with the $K = 2$ state by 0.92.
Therefore, the second $2^+$ state has the component of 
$K = 0$ by 8$\%$, and it can be considered
as an indication of a triaxial deformation. 

This triaxiality of the $2^+$ states is reflected in
the electro-magnetic transition rate, and B(E2) values 
are summarized in Table II.

\noindent
\begin{center}
--------------\\
Table II \\
--------------\\
\end{center}
The E2 transition between the first $2^+$ state and the
ground state 
is calculated as B(E2: $2^+_1$ $\to$ $0^+_1$) = 11.8 $e^2$fm$^4$,
which agrees with the experimental value of 10.04$\pm$1.2 $e^2$fm$^4$.
Furthermore, the interband transition is calculated:
B(E2: $2^+_1$ $\to$ $2^+_2$) = 3.99 $e^2$fm$^4$. 
If the system is axially symmetric, this transition 
between different $K$-values is more suppressed.
Therefore, the value indicates the $2^+$ states have
component of a triaxial deformation.

Using Davydov-Filippov model\cite{Davydov}, we can estimate the
degree of the triaxiality as a function of the $\gamma$ angle.
The ratios
{${\rm B(E2: } \ 2^+_2 \to 0^+_1) \over {\rm B(E2:} \ 2^+_1 \to 0^+_1)$}
and 
{${\rm B(E2: } \ 2^+_2 \to 2^+_1) \over {\rm B(E2:} \ 2^+_1 \to 0^+_1)$}
are given in Davydov-Filippov model as follows:
\begin{equation}
{{\rm B(E2: } \ 2^+_2 \to 0^+_1) \over {\rm B(E2:} \ 2^+_1 \to 0^+_1)}
= { 1-{3-2\sin^2(3\gamma) \over \sqrt{9-8\sin^2(3\gamma)}}
\over 1+{3-2\sin^2(3\gamma) \over \sqrt{9-8\sin^2(3\gamma)}} },
\end{equation}
\begin{equation}
{{\rm B(E2: } \ 2^+_2 \to 2^+_1) \over {\rm B(E2:} \ 2^+_1 \to 0^+_1)}
= { {20 \over 7} { \sin^2(3\gamma) \over 9-8\sin^2(3\gamma)}
\over 1+{3-2\sin^2(3\gamma) \over \sqrt{9-8\sin^2(3\gamma)}} }.
\end{equation}
These values are compared with our calculation in Fig. 3.

\noindent
\begin{center}
--------------\\
Fig. 3 \\
--------------\\
\end{center}
The solid line in Fig. 3 shows the
ratio
{${\rm B(E2: } \ 2^+_2 \to 2^+_1) \over {\rm B(E2:} \ 2^+_1 \to 0^+_1)$}
calculated with the Davidov-Filippov model.
The ratio becomes 0.34 in our calculation, and it has a
crossing point with
Davydov-Filippov model around $\gamma = 19^o$.
The dotted line in Fig. 3 shows the ratio
{${\rm B(E2: } \ 2^+_2 \to 0^+) \over {\rm B(E2:} \ 2^+_1 \to 0^+_1)$}
calculated with the Davidov-Filippov model,
and it crosses with our result of 0.059
around $\gamma =$ 17$^o$ and 22$^o$.
These results strongly suggest that $^{10}$Be 
has a triaxial deformation of $\gamma = 15^o \sim 20^o$.
Although the $\alpha$-$\alpha$ core is of axial symmetry
and electric charge are only in the $\alpha$'s,
the recoil effect gives rise to a change from the axial symmetry 
to the triaxial shape.

The $\gamma$-value can be deduced also in terms of
$Q$-invariant model.
In Ref. \cite{Werner}, the $\gamma$-value
is related to the reduced matrix elements of
$Q$ operator:
\begin{equation}
q_2 = \sum_i \langle 0^+_1 || Q || 2^+_i \rangle 
             \langle 2^+_i || Q || 0^+_1 \rangle,
\end{equation}
\begin{equation}
q_3 = {\sqrt{7 \over 10}}\sum_{i,j} \langle 0^+_1 || Q || 2^+_i \rangle 
                                   \langle 2^+_i || Q || 2^+_j \rangle
                                   \langle 2^+_j || Q || 0^+_1 \rangle,
\end{equation}
\begin{equation}
{q_3 \over {q_2}^{3 \over 2}} = \cos 3\gamma.
\end{equation}
Using these relations and taking the sum over indices $i$ and $j$
up to 2, our calculated results correspond to $\gamma = 20.1^o$.

We discuss the electro-magnetic transition between the two
$2^+$ states ($K$-mixing effect between the two $2^+$ states)
by artificially changing the strength parameter $V^{ls}_0$ 
in Eq. (15) for the
spin-orbit term. The Majorana parameter $M$ for the central
term is simultaneously changed to keep the calculated binding energy constant.
In Table III,  
the B(E2: $2^+_1$ $\to$ $2^+_2$) values are listed 
together with the calculated energies and magnetic dipole moments. 
Using the original interaction, 
the B(E2: $2^+_1$ $\to$ $2^+_2$) value 
is predicted as 3.99 $e^2$fm$^4$.
  
\noindent
\begin{center}
--------------\\
Table III \\
--------------\\
\end{center}
When we increase the $V^{ls}_0$ value, 
the excitation energy of the second $2^+$ state becomes higher,
since one of the valence neutrons 
repulsively feels the spin-orbit interaction. 
The original interaction ($V^{ls}_0 = 2000$ MeV and $M = 0.6$)
gives $E_x = 5.70$ MeV for the $2^+_2$ state,
but the interaction with $V^{ls}_0 = 2500$ MeV and $M = 0.61$
gives $E_x = 6.82$ MeV.
The B(E2: $2^+_1$ $\to$ $2^+_2$) value then decreases
from 3.99 $e^2$fm$^4$ to 2.20 $e^2$fm$^4$.
Therefore, it is considered that 
with increasing $LS$ strength,
the $K$ quantum number of each $2^+$ state
approaches to a good number,
where the E2 transition between the two $2^+$ states is suppressed.
On the other hand, when the spin-orbit interaction 
becomes weaker, the transition rapidly increases.
The $V^{ls}_0$ value of 1500 MeV gives 
the B(E2: $2^+_1$ $\to$ $2^+_2$) values of 9.53 $e^2$fm$^4$,
and when we adopt $V^{ls}_0 = 1000$ MeV, the value becomes 
17.53 $e^2$fm$^4$.
Here, the orbits of the valence neutrons deviate from ones
with good $K$ quantum
numbers, and a triaxial $\alpha$+$\alpha$+di-neutron clustering
configuration where the two valence neutrons
are strongly correlated becomes important.

We can intuitively interpret this behavior as follows:
when the spin-orbit is weak enough, the two valence neutrons
form di-neutron, in which
the attractive interaction between them strongly contributes.
However, when the spin-orbit interaction significantly acts
and becomes more important, the di-neutron is broken 
and each valence neutron rotates around 
the core in opposite direction with definite $K$-values.
Note that
the spin-orbit interaction does not act to di-neutron with $S = 0$.
This is close to $jj$-coupling 
picture and axial symmetry of the system is restored.

This situation can be interpreted also from nuclear $SU_3$ model.
At the $SU_3$ limit (parameters $a$, $b$, and $d$ $\to$ 0),
$\Phi((3/2^-)^2)$ (dominant configuration of $2^+_1$) 
and $\Phi(3/2^-\ 1/2^-)$ (dominant configuration of $2^+_2$) 
correspond to linear combinations of two $SU_3$ configurations:
$(n_x, n_y, n_z)$ = (2,0,4), $(\lambda, \mu)$ = (2,2)
and $(n_x, n_y, n_z)$ = (1,1,4), $(\lambda,\mu)$ = (3,0)
($n_z = 4$ is due to the $\alpha$-clusters along the $z$-axis).
This is because,
the valence neutrons in $3/2^-$ and $1/2^-$ 
are expressed as linear combinations of $p$-orbitals along
the $x$- and $y$-direction.
Here, the spin-orbit interaction determines the mixing ratio
of the two $SU_3$ representations.
When it is strong enough,
since the $jj$-coupling picture works well, 
these two configurations are mixed to the
almost same amount. However, when the
spin-orbit interaction is weakened, the binding-energy gain due to
the mixing amplitudes changes.
Here, the $(\lambda, \mu) = (2,2)$ configuration,
which is a di-neutron configuration, becomes much more important 
than the $(\lambda, \mu) = (3,1)$ configuration.
This means that a triaxial deformation is induced when the 
spin-orbit interaction becomes weaker.

Next, we examine the magnetic dipole moment as a probe to
determine the strength of the spin-orbit interaction. 
In Table III, the magnetic dipole moments of these two $2^+$ states
are listed, and we predict $\mu = 0.72 \mu_N$ and $0.48 \mu_N$
for the first and the second $2^+$ states, respectively.  
Unfortunately, the dependence of the total magnetic moment of
these states with respect to the spin-orbit strength is rather 
monotonic. However, each component has a significant dependence.
In Table III, proton-orbital part and neutron-spin part
are also listed (neutron-orbital part is of course zero,
and proton-spin part also becomes zero due to the 
assumption of the $\alpha$-clusters).  

As for the proton-orbital part, 
using the original interaction $(V^{ls}_0 = 2000$ MeV, $M = 0.6$),
$2^+_1$ is calculated to have a larger value than
$2^+_2$: the $2^+_1$ state has 1.13 $\mu_N$ and the $2^+_2$ state
has 0.46 $\mu_N$.
The $2^+_1$ state is mainly of $K = 0$, 
and therefore the rotation of the $\alpha$-$\alpha$ core is 
the main source of the angular momentum ($J = 2$). 
However, in the case of $2^+_2$,
since the valence neutrons already have $K = 2$,
the rotation of $\alpha$-$\alpha$
is not necessary to construct $J = 2$, and hence the proton-orbital
part is smaller than $2^+_1$.
These values depend on the strength of the spin-orbit interaction.
With decreasing spin-orbit strength, the difference 
of proton-orbital part between these two states becomes smaller 
due to the increase of the 
$K$-mixing effect between the two $2^+$ configurations,
which we discussed.
When the strength of the spin-orbit interaction 
is $V^{ls}_0 = 1500$ MeV,
$2^+_1$ has 1.04 $\mu_N$ and $2^+_2$ has 0.65 $\mu_N$.
As for the neutron-spin part, 
in $V^{ls}_0 = 2000$ MeV case (original interaction),
$2^+_1$ has $-0.41$ $\mu_N$ and $2^+_2$ has $-0.04$ $\mu_N$.
This result suggests that the $2^+_1$ state has both 
spin-singlet and spin-triplet components
of the valence neutrons. The $2^+_2$ state is dominated by
the spin-singlet component.
In $2^+_2$, one of the valence neutrons 
repulsively feels the spin-orbit interaction,
and when the two valence neutrons construct spin-singlet,
it helps to reduce the contribution of the repulsive 
spin-orbit interaction.
This is the main reason for the smaller neutron-spin part in $2^+_2$.
On the other hand, if we use weaker spin-orbit strength,
the difference of neutron-spin part between the two $2^+$ states
becomes much smaller.
When the strength of the spin-orbit interaction 
is $V^{ls}_0 = 1000$ MeV,
$2^+_1$ has $-0.13$ $\mu_N$ and $2^+_2$ has $-0.12$ $\mu_N$.

Finally, we comment 
that the $K$-mixing effect is also important for the $4^+$ states. 
The $4^+$ state with $K = 0$
at $-47.6$ MeV in Fig. 2 becomes $-50.0$ MeV after the $K$-mixing, 
and the $4^+$ state with $K = 2$ at $-46.1$ MeV
is pushed up to $-43.5$ MeV.
The first $4^+$ state has the squared overlap of 0.78 with the $K=0$
while 0.63 
with the $K = 2$ state (here, $K = 0$ and $K = 2$ are not orthogonal).
Therefore, the first $4^+$ state has almost equal contribution 
of the $K=0$ and the $K = 2$ basis states.
This is because, in the case of $4^+$, 
the angular momentum vector 
with $K = 0$ and $K = 2$ are not necessary to be spatially orthogonal,
different from the case of $2^+$.
Therefore, 
the $K$-mixing for the $4^+$ state of $^{10}$Be is very strong,
and the electro-magnetic transition probability
among the yrast band deviates from a simple rigid-body picture.
The B(E2: $4^+_1 \to 2^+_1$) value is calculated to be 11.1
$e^2$fm$^4$,
even smaller than
the B(E2: $2^+_1 \to 0^+_1$) value of 11.8 $e^2$fm$^4$, and 
the ${{\rm B(E2: } \  4^+_1 \to 2^+_1) \over {\rm B(E2:} \ 2^+_1 \to 0^+_1)}$
ratio of 0.94 much deviates from the rigid-body limit of 1.43.
If we restrict the model space to $K=0$, then the 
B(E2: $4^+_1 \to 2^+_1$) becomes 14.7 $e^2$fm$^4$,
and the ratio 
{${\rm B(E2: } \ 4^+_1 \to 2^+_1) \over {\rm B(E2:} \ 2^+_1 \to 0^+_1)$}
of 1.18 becomes closer to 1.43.
In nuclear $SU_3$ model, the $4^+$ states with $K = 0$
$((\lambda, \mu) = (2,2))$
and $K = 2$ $((\lambda, \mu) = (3,1))$
become the same representation. This character partially remains 
as a strong $K$-mixing effect in the present MO model.

%%%%%%%%%%%%%%%%%%%%%%%%%%%%%%%%%%%%%%%%%%%%%%%%%%%%%%%%%%%%%%%%%
\section{Summary and conclusion}
We have applied
the $\alpha$+$\alpha$+$n$+$n$ model to $^{10}$Be
and discussed the triaxial deformation of this nucleus.
The orbits for the valence neutrons have been introduced 
based on the molecular orbit (MO) model.
In the present model,
the spatially extended motion of the valence neutrons 
around the $\alpha$-clusters are described
by linear combinations of Gaussians, and 
the centers of the Gaussians are variationally determined.

The calculated energy levels show the appearance of 
two band structures, which have dominantly
$K=0$ and $K=2$ components, when
the two valence neutrons occupy $K^\pi = 3/2^-$ or $K^\pi = 1/2^-$
orbits.
The first $2^+$ state has the squared overlap with the $K=0$
state by 0.96, and the second $2^+$ state has the squared overlap
with the $K=2$ state by 0.92.
Since the second $2^+$ state has the component of 
$K=0$ by 8$\%$, 
the electro-magnetic transition
from the $2^+_1$ state (mainly $K = 0$) to 
the $2^+_2$ state (mainly $K = 2$) is allowed
(3.99 $e^2$fm$^4$). 
Using Davydov-Filippov model\cite{Davydov}, we estimated the
triaxiality for the $2^+$ state.
The ratio
{${\rm B(E2: } \ 2^+_2 \to 2_1^+) \over {\rm B(E2:} \ 2^+_1 \to 0^+_1)$}
calculated with the Davidov-Filippov model and
our model coincide around $\gamma = 19^o$,
and B(E2) ratio
{${\rm B(E2: } \ 2^+_2 \to 0^+) \over {\rm B(E2:} \ 2^+_1 \to 0^+_1)$}
indicates $\gamma =$ 17$^o$ $\sim$ 22$^o$.
Originally, the $\alpha$-$\alpha$ core is introduced to be of axial symmetry.
However, because of the recoil effect of the valence neutrons,
the charge distribution deviates from the axial symmetry, 
and the system becomes triaxial.
We also discussed the triaxial deformation
by artificially weakening the spin-orbit interaction.
With decreasing spin-orbit interaction,
the orbits of the valence neutrons deviate from the $jj$-coupling
limit, and the di-neutron configuration becomes important.
Here, the system becomes 3 body-like and
the ${\rm B(E2: } \ 2^+_2 \to 2^+_1)$ value drastically increases. 

Similar di-neutron component
is discussed in weakly bound systems with the so-called halo structure.
For example, in $^6$He, in addition to the shell model-like space, 
the model space of di-neutron+$^4$He has been shown 
to be important \cite{Aoyama}.
This means that a locally-correlated di-neutron wave function
is important for the description of the valence neutrons
with small binding energy and spatially extended distribution.
It is consistent with our discussion 
on the effect of varying  the strength of the spin-orbit interaction.
Namely, when the valence neutrons with low-binding energies
have halo structure, the contribution of the spin-orbit
interaction between the core and the valence neutrons becomes weak.
Here, the valence neutrons 
construct a di-neutron pair with spin singlet,
by which they can increase the spatial overlap between them
and the contribution of the attractive interaction.
Therefore, it is very challenging 
to explore the triaxial deformations in $^{12}$Be and $^{14}$Be,
which have weakly bound neutrons and also deformed cores.

\vspace*{1cm}

The authors thank members of Nuclear Theory group 
in University of Tokyo and RI beam science laboratory
in RIKEN for discussions and encouragements.
They also thank Dr. I. Kumagai-Fuse for her help.
One of the authors (N.I) thanks 
Prof. R. Lovas, Prof. W. von Oertzen,
Prof. H. Horiuchi, Prof. K. Kat\=o, and Dr. Y. Kanada-En'yo
for fruitful discussions. 
This work is supported in part by Grant-in-Aid 
for Scientific Research (13740145) from the Ministry of Education,
Science and Culture.

%%%%%%%%%%%%%%%%%%%%%%%%%%%%%%%%%%%%%%%%%%%%%%%%%%%%%%%%%%%%%%%%%%
%    References

%%%%%%%%%%%%%%%%%%%%%%%%%%%%%%%%%%%%%%%%%%%%%%%%%%%%%%%%%%%%%%%%%%
%    Figure and Table
%******************************************************************8
% Fig. 1
%******************************************************************8
\begin{figure}
\GETPS{4}{fig1.ps}
\caption{
The intrinsic density of $\Phi((3/2^-)^2)$ 
((a): $xz$-plane, (b): $xy$-plane), 
where the $\alpha$-$\alpha$ distance is chosen to be optimal 3 fm.
%The contour line of 0.01 corresponds to $\beta = 0.5 $ amd $\gamma = 11^o$.
}
\end{figure}
%******************************************************************8
% Fig. 2
%******************************************************************8
\begin{figure}
\GETPS{4}{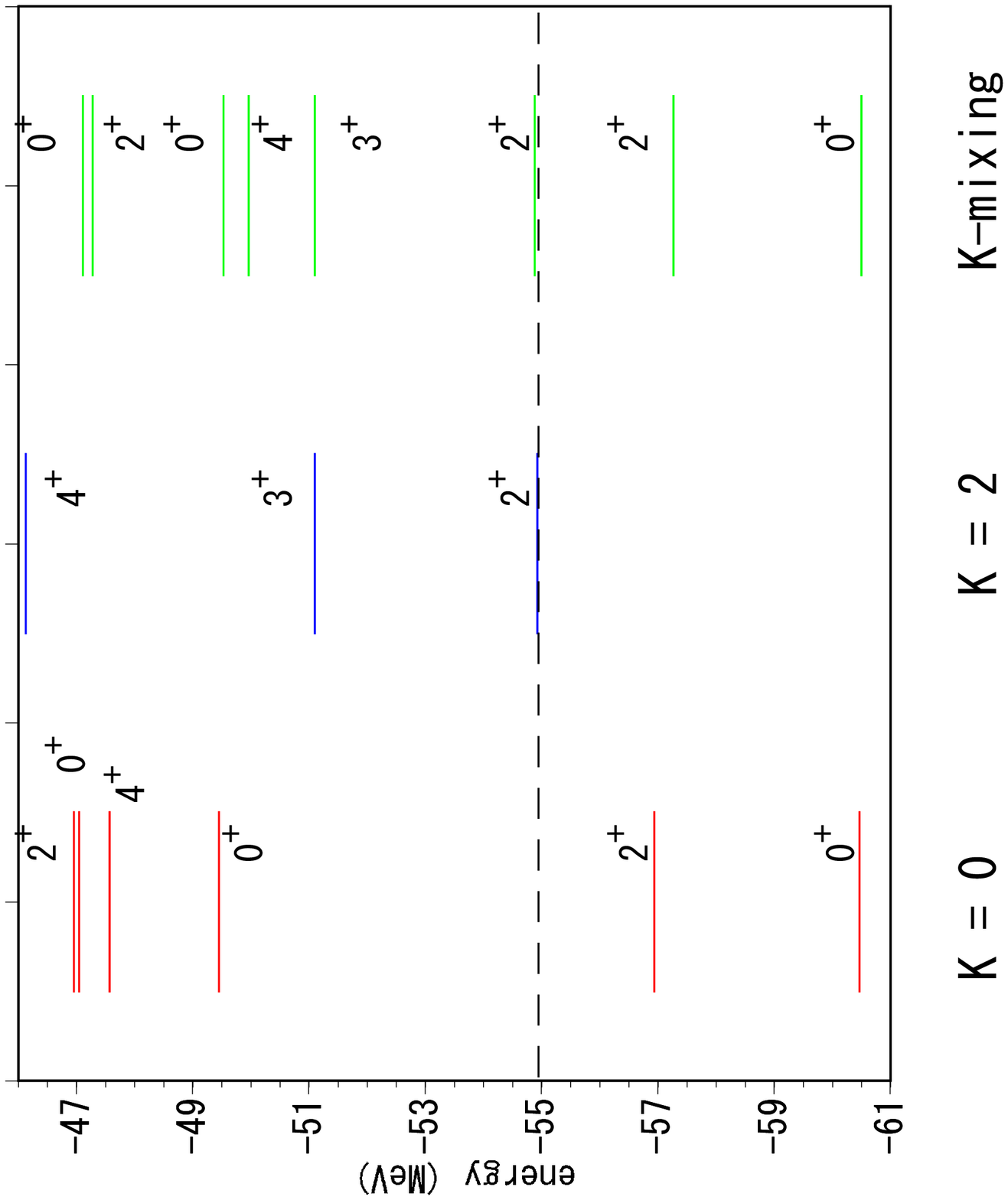}
%\centerline{\psfig{figure=schem.ps,height=15cm}}
\caption{
The calculated energy levels of $^{10}$Be.
The left and the middle column 
show the levels calculated with the constraints of $K = 0$ and $K = 2$.
The right column shows the calculated energy levels after $K$-mixing.
The calculated $\alpha$+$\alpha$+$n$+$n$ threshold energy ($-55.0$ MeV)
is shown as the dashed line.
Experimentally, the $2^+_1$ and the $2^+_2$ states
are observed at $E_x = 3.3$ MeV and $E_x = 5.9$ MeV, respectively.
} 
\end{figure}
%******************************************************************8
% Fig. 3
%******************************************************************8
\begin{figure}
%\centerline{\psfig{figure=schem.ps,height=15cm}}
\GETPS{4}{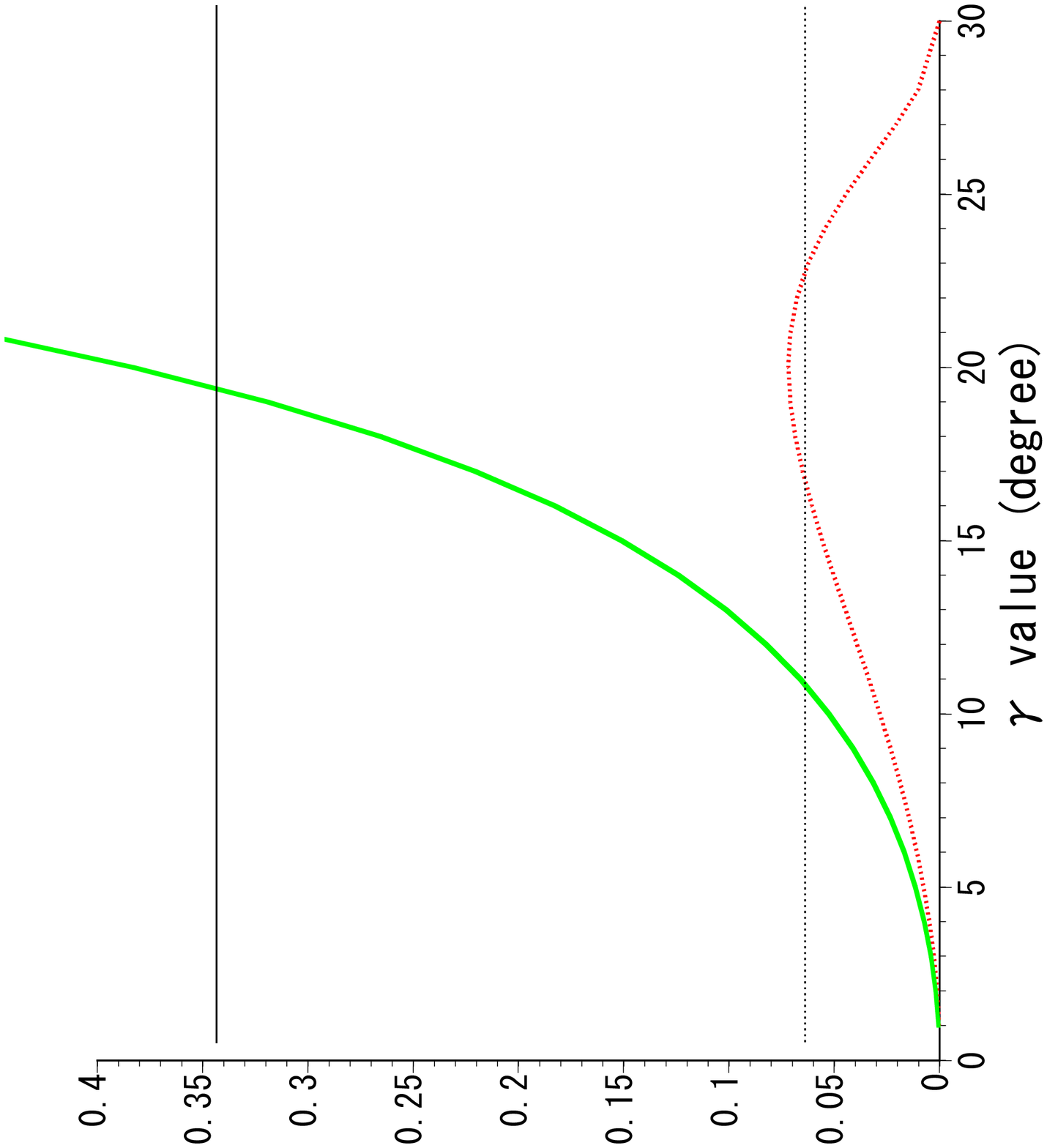}
\caption{
B(E2) ratios
{${\rm B(E2: } \ 2^+_2 \to 2^+_1) \over {\rm B(E2:} \ 2^+_1 \to 0^+_1)$}
(solid line) and
{${\rm B(E2: } \ 2^+_2 \to 0^+_1) \over {\rm B(E2:} \ 2^+_1 \to 0^+_1)$}
(dotted line), as a function of $\gamma$ (degree).
Our results of 0.34 and 0.059 cross with Davydov-Filippov
model around 15$^o \sim 20^o$.
}
\end{figure}
%******************************************************************8
%     Table I
%******************************************************************8
\begin{table}
\caption{
The values of $a$, $b$ and the intrinsic energy
as a function of the $\alpha$-$\alpha$ distance. 
Parameter $a$ for the spin-up and the spin-down
valence neutrons $a_1$ and $a_2$, respectively,
and parameter $b$ for them is $b_1$ and $b_2$.
The wave function of $^{10}$Be 
is $\Phi((3/2^-)^2)$ (upper panel), 
$\Phi((1/2^-)^2)$ (middle panel),
and $\Phi(3/2^-,\ 1/2^-)$ (lower panel).
}
\begin{center}
\begin{tabular}{|c|c|c|c|}
$\alpha$-$\alpha$ (fm) & $a_1 = a_2$ (fm) & 
$b$ (fm) & intrinsic (MeV)\\
\hline
2 & 1.09 & 0.99 & $-52.09$  \\
3 & 1.39 & 1.08 & $-50.91$  \\
4 & 1.71 & 1.35 & $-44.64$  \\
5 & 2.12 & 1.53 & $-38.04$  \\
\end{tabular}
\begin{tabular}{|c|c|c|c|}
\hline
$\alpha$-$\alpha$ (fm) & $a_1 = a_2$ (fm) 
& $b$ (fm) & intrinsic (MeV) \\
\hline
2 & 1.33 & 2.05 & $-38.06$  \\
3 & 1.49 & 2.11 & $-39.30$  \\
4 & 1.63 & 2.21 & $-35.42$  \\
5 & 1.89 & 2.35 & $-30.34$  \\
\end{tabular}
\begin{tabular}{|c|c|c|c|c|c|}
\hline
$\alpha$-$\alpha$ (fm) & $a_1$ (fm) & $a_2$ (fm) 
& $b_1$ (fm) & $b_2$ (fm) & intrinsic (MeV) \\
\hline
2 & 1.12 & 1.18 & 0.83 & 1.87 & $-50.20$ \\
3 & 1.33 & 1.37 & 1.12 & 1.94 & $-49.81$ \\
4 & 1.66 & 1.59 & 1.40 & 2.09 & $-44.16$ \\
5 & 2.05 & 2.36 & 1.59 & 2.23 & $-37.79$ \\
\end{tabular}
\end{center}
\end{table}
%******************************************************************8
%     Table II
%******************************************************************8
\begin{table}
\caption{The electro-magnetic transition probability (B(E2))
in $^{10}$Be after the $K$-mixing. All units are $e^2$fm$^4$.}

\begin{center}
\begin{tabular}{|c|c|}
B(E2: $2^+_1$ $\to$ $0^+_1$ ) & 11.8 (Exp.\ 10.04$\pm$1.2) \\
B(E2: $2^+_2$ $\to$ $0^+_1$ ) & 0.70 \\
B(E2: $2^+_1$ $\to$ $2^+_2$ ) & 3.99 \\
B(E2: $4^+_2$ $\to$ $2^+_1$ ) & 11.1 \\
\end{tabular}
\end{center}
\end{table}

%******************************************************************8
%     Table III
%******************************************************************8
\begin{table}
\caption{The electro-magnetic transition probability (B(E2))
from first $2^+$ state to second $2^+$ calculated by changing
the strength of the spin-orbit interaction $(V^{ls}_0)$.
The Majorana parameter $(M$) for the central interaction
is also changed to keep the calculated binding energy constant.
The original interaction is $V^{ls}_0 = 2000$ MeV and $M = 0.6$.
Magnetic dipole moment is also predicted.
$\mu(pl)$ and $\mu(ns)$ represent proton-orbital part
and neutron-spin part, respectively.}

\begin{center}
\begin{tabular}{|c|c|c|c|c|}
$V^{ls}_0$ (MeV) & 1000 & 1500 & 2000  & 2500 \\
$M$ & 0.58 & 0.59 & 0.60 & 0.61 \\
\hline
$0^+_1$ (MeV) & $-60.48$ & $-60.38$ & $-60.51$ & $-60.75$ \\
$2^+_1$ (MeV) & $-57.69$ & $-57.30$ & $-57.26$ & $-57.33$ \\
$2^+_2$ (MeV) & $-56.71$ & $-55.78$ & $-54.88$ & $-53.93$ \\
\hline
B(E2: $2^+_1$ $\to$ $2^+_2$) ($e^2$fm$^4$) & 17.53 & 9.53 & 3.99 & 2.20 \\
\hline
$\mu$ ($2^+_1$) ($\mu_N$) & 0.57 & 0.73 & 0.72 & 0.69 \\
($\mu(pl)$, $\mu(ns))$ & (0.70, $-0.13$) & (1.04, $-0.30$) & 
(1.13, $-0.41$) & (1.16, $-0.48$)  \\
\hline
$\mu$ ($2^+_2$) ($\mu_N$) & 0.91 & 0.59 & 0.48 & 0.43 \\
($\mu(pl)$, $\mu(ns))$ & (1.03, $-0.12$) & (0.65, $-0.06$) & 
(0.51, $-0.04$) & (0.46, $-0.03$)  \\
\end{tabular}
\end{center}
\end{table}


\newpage
\begin{thebibliography}{99}
\bibitem{Iwasaki1} H. Iwasaki $et\ al.$,
Phys. Lett. {\bf B}481, 7 (2000).
\bibitem{Iwasaki2} H. Iwasaki $et\ al.$,
Phys. Lett. {\bf B}491, 8 (2000).
\bibitem{Navin} A. Navin $et\ al.$,
Phys. Rev. Lett. {\bf 85}, 266 (2000).
\bibitem{Kor}
A. A. Korscheninnikov $et\ al.$ Phys. Lett. {\bf B}343, 53 (1995).
\bibitem{RB}
N. Soi\' c $et$ $al$, Eurohys. Lett, {34(1)}, 7 (1996).
\bibitem{Freer1}
M. Freer $et\ al.$,  Phys. Rev. Lett. {\bf 82}, 1383 (1999).
\bibitem{Freer2}
M. Freer $et\ al.$,  Phys. Rev. C{\bf 63}, 034301 (2001).
\bibitem{Ita}
N. Itagaki and S. Okabe, Phys. Rev. C {\bf 61} 044306, (2000).
\bibitem{Ita2}
N. Itagaki, S. Okabe, and I. Ikeda, Phys. Rev. C{\bf 62}, 034301 (2000).
\bibitem{Ita3}
N. Itagaki, S. Okabe, K. Ikeda, and I. Tanihata,
Phys. Rev. C {\bf 64} 014301, (2001).
\bibitem{Harvey}
M. Harvey and F. C. Khanna, Nuclear Spectroscopy and Reactions, Part D,
Models of Light Nuclei, Ed. by J. Cerny, (Academic Press Inc., New York, 1975),
69.
\bibitem{Davydov}
A.S. Davydov and G.F. Filippov, Nucl. Phys. 8, 237 (1958).
\bibitem{Curtis}
N. Curtis $et\ al.$, submitted to Phys. Rev. C
\bibitem{Abe}
Y. Abe, J. Hiura, and H. Tanaka, Prog. Theor. Phys. {\bf 49}, 800 (1973).
\bibitem{Okabe-S}
H. Furutani, H. Kanada, T. Kaneko, S. Nagata, H. Nishioka, S. Okabe
S. Saito, T. Sakuda and M. Seya,
Prog. Theor. Phys. Suppl. {68}, 193 (1980).
\bibitem{Seya}
M. Seya, M. Kohno and S. Nagata,
Prog. Theor. Phys. {65}, 204 (1981).
\bibitem{EnyoBe10} Y. Kanada-En'yo, H. Horiuchi and A.Dot{\'e}, 
Phys. Rev. C{\bf 60}, 064304 (1999).
\bibitem{Ogawa} Y. Ogawa, K. Arai, Y. Suzuki, and K. Varga,
Nucl. Phys. {\bf A}673, 122 (2000).
\bibitem{Enyo95}
Y. Kanada-En'yo and H. Horiuchi, 
Prog. Theor. Phys. {93}, 115 (1995).
\bibitem{Enyo95b}
Y. Kanada-En'yo, H. Horiuchi and A. Ono, 
Phys. Rev. C{\bf 52}, 628 (1995).
\bibitem{Enyo95c}
Y. Kanada-En'yo and H. Horiuchi 
Phys. Rev. C{\bf 52}, 647 (1995).
\bibitem{AMD} A. Dot{\'e}, H. Horiuchi and Y. Kanada-En'yo,
Phys. Rev. C{\bf 56}, 1844 (1997).
\bibitem{Ono92}
A. Ono, H. Horiuchi, T. Maruyama and A. Ohnishi,\\
Prog. Theor. Phys. {87} 1185, (1992);
Phys. Rev. Lett. {68} 2898, (1992).
\bibitem{VolkovInt}
A.B. Volkov Nucl. Phys. {74}, 33 (1965).
\bibitem{G3RS} N. Yamaguchi, T. Kasahara, S. Nagata and Y. Akaishi,
Prog. Theor. Phys. 62 1018 (1979)
\bibitem{Okabe79}
S. Okabe and Y. Abe, Prog. Theor. Phys. {61}, 1049 (1979).
\bibitem{Werner}
V. Werner, N. Pietralla, P. von Brentano, R.F. Casten, and 
R.V. Jolos, Phys. Rev. C{\bf 61}, 021301(R) (2000).
\bibitem{Aoyama}
S. Aoyama, S. Mukai, K. Kat\=o, and K. Ikeda,
Prog. Theor. Phys. {\bf 93}, 99 (1995).
%\bibitem{EnyoBe10} Y. Kanada-En'yo, H. Horiuchi and A. Dot{\'e}, 
%Phys. Rev. C {\bf 60}, 064304 (1999).
\end{thebibliography}
\end{document}